\documentclass[twocolumn,showpacs,preprintnumbers,amsmath,amssymb,nofootinbib,superscriptaddress]{revtex4-1}

\usepackage[colorlinks,pdfstartview=FitH]{hyperref}
\hypersetup{linkcolor=blue,citecolor=blue,filecolor=black,urlcolor=blue}

\usepackage{amsmath,amssymb,bm}
\usepackage{graphicx}
\usepackage{amsfonts}
\usepackage{mathtools}
\usepackage{color}

\begin{document}

\title{Interpretation of XENON1T excess with MeV boosted dark matter}

\author{Lian-Bao Jia}
\email{jialb@mail.nankai.edu.cn}
\affiliation{
School of Science, Southwest University of Science and Technology, Mianyang 621010, China
}
\author{Tong Li}
\email{litong@nankai.edu.cn}
\affiliation{
School of Physics, Nankai University, Tianjin 300071, China
}

\begin{abstract}
The XENON1T excess of keV electron recoil events may be induced by the scattering of electrons and long-lived particles with MeV mass and high-speed. We consider a tangible model composed of two scalar MeV dark matter (DM) particles $S_A$ and $S_B$ to interpret the XENON1T keV excess via boosted $S_B$. A small mass splitting $m_{S_A}-m_{S_B}>0$ is introduced and the boosted $S_B$ can be produced by the dark annihilation process of $S_A S_A^\dagger \to \phi \to S_B S_B^\dagger$ via a resonant scalar $\phi$. The $S_B-$electron scattering is intermediated by a vector boson $X$. Although the constraints from BBN, CMB and low-energy experiments set the $X-$mediated $S_B-$electron scattering cross section to be $\lesssim 10^{-35} \mathrm{cm}^2$, the MeV scale DM with a resonance enhanced dark annihilation today can still provide enough boosted $S_B$ and induce the XENON1T keV excess. The relic density of $S_B$ is significantly reduced by the $s$-wave process of $S_B S_B^\dagger \to X X$ which is allowed by the constraints from CMB and 21-cm absorption. A very small relic fraction of $S_B$ is compatible with the stringent bounds on un-boosted $S_B$-electron scattering in DM direct detection and the $S_A$-electron scattering is also allowed.
\end{abstract}

%\begin{document}

\maketitle
%\flushbottom
%\newpage

%%%%%%%%%%%%%%%%%%%%%%%%%%%%%%%%%%%%%%%%%%%%%%%
\section{Introduction}
\label{sec:Intro}
%%%%%%%%%%%%%%%%%%%%%%%%%%%%%%%%%%%%%%%%%%%%%%%

The existence of Dark Matter (DM) has been established by substantial cosmological and astronomical observations. However, the microscopic properties of DM beyond the Standard Model (SM) are largely unknown. Recently, the XENON collaboration reported an excess of electronic recoil events with the energy around 2-3 keV \cite{XENON:2020rca}, and the event distribution has a broad spectrum for the excess. They collected low energy electron recoil data from the XENON1T experiment with an exposure of 0.65 tonne-years and analyzed various backgrounds for the excess events.
Although a small tritium background fits the excess data well~\cite{XENON:2020rca}, the bosonic DM can also provide a plausible source for the peak-like excess.

The excess of electron recoil events may be induced by new long-lived particle scattering with electrons in detector. The lifetime of the new particle needs to be long enough to reach the detector on Earth after its production and it has an appreciable interaction with electron. The mass of the new long-lived particle should be $\gtrsim$ MeV and the velocity is at the level of $\mathcal{O}(0.1)$c~\cite{Kannike:2020agf}. Meanwhile, it should be compatible with the structure formation of the universe and the constraints from the big bang nucleosynthesis (BBN) as well as the cosmic microwave background (CMB). Thus, some exotic mechanism is needed to produce a number of long-lived particles with a high speed. A plausible scenario for the electron excess events is boosted DM produced in the present universe via dark sector annihilation or decay~\cite{Kannike:2020agf,Fornal:2020npv,Du:2020ybt,DelleRose:2020pbh,Alhazmi:2020fju,Davoudiasl:2020ypv,VanDong:2020bkg,Cao:2020oxq}. Meanwhile, a fraction of the un-boosted DM with a regular velocity distribution (with a velocity of $\sim 10^{-3}~c$ in the Galaxy) can also be present today and is detectable via the scattering off electron. The direct detection experiments would thus set stringent bounds on the un-boosted DM-electron scattering for the recoil energy of a few eV~\cite{XENON10:2011prx,Essig:2017kqs,DarkSide:2018ppu,XENON:2019gfn,SENSEI:2020dpa}. These bounds need to be evaded when interpreting the XENON1T keV excess.

To interpret the XENON1T excess via the scattering between electron and boosted DM, the nature of the intermediating particle and the interaction becomes a key question. For a long-lived light mediator with keV$\sim$MeV mass, the BBN and CMB would place stringent constraints \cite{Slatyer:2016qyl,Kawasaki:2020qxm}. If the new mediator has a short lifetime, given the constraint from BBN, its mass should be $\gtrsim$ 10 MeV and its lifetime is much shorter than a second~\cite{Ho:2012ug,Boehm:2013jpa,Jia:2016uxs,Berlin:2018sjs}. In addition, the constraints from low-energy experiments, such as NA48/2 \cite{NA482:2015wmo} and NA64~\cite{NA64:2019auh}, should be considered as well. On the other hand, the new mediator and interaction may leave some traces in anomalous processes, e.g., a new vector boson about 17 MeV \cite{Feng:2016jff,Feng:2020mbt} predominantly decaying into $e^+ e^-$ was suggested by two anomalous transitions of $^8$Be~\cite{Krasznahorkay:2015iga} and $^4$He~\cite{Krasznahorkay:2019lyl}. Here we consider a light vector boson $X$ in general which mainly decays into $e^+ e^-$ and intermediates the scattering between electron and the boosted DM. Considering the constraints from low-energy experiments~\cite{NA482:2015wmo,NA64:2019auh}, when $14~{\rm MeV}\lesssim m_X\lesssim 30$ MeV as shown in Fig.~5 of Ref.~\cite{NA64:2019auh}, a part of the parameter $\epsilon_e$ (the $X-$electron coupling is parameterized as $\epsilon_e e$) in the range of $10^{-4} \lesssim \epsilon_e \lesssim 10^{-3}$ is still allowed by the experiments.~\footnote{Note that the rapid fluctuations in the NA48/2 limit~\cite{NA482:2015wmo} cause some uncertainty of the NA48/2 limit. Here we take the NA48/2 limit in a smooth way as shown in Fig.~5 of Ref.~\cite{NA64:2019auh} instead of the rapid fluctuations.} For instance, the range of $5\times 10^{-4} \lesssim \epsilon_e \lesssim 10^{-3}$ is allowed for $m_X \sim 20$ MeV of our interest.

In this paper we introduce two complex scalar DM particles $S_A$ and $S_B$ to interpret the XENON1T excess with a light vector mediator $X$. The DM particles $S_A$ and $S_B$ are under possible dark symmetry in the hidden sector with $m_{S_A}\simeq m_{S_B}$, and some dark sector numbers are carried by both $S_A$ and $S_B$ to keep the stability of DM. The DM particle $S_B$ is dark charged and $S_A$ is neutral with a small mass splitting $m_{S_A}-m_{S_B}>0$ which can be introduced from radiative corrections or substructures. The pair of $S_B S_B^\dagger$ can be produced via the dark annihilation process of $S_A S_A^\dagger$ $\to S_B S_B^\dagger$ mediated by a new scalar $\phi$. The present dark annihilation can thus provide a source of boosted $S_B$. The scattering between electron and the boosted $S_B$ mediated by $X$ boson may explain the keV electron excess observed by XENON1T. Besides the boosted $S_B$ accounting for the XENON1T keV excess, there would be a fraction of un-boosted $S_B$ around the Earth. The relic abundance of $S_B$ could be significantly reduced by the transition of $S_B S_B^\dagger \to $ $X X$. Thus, it will be compatible with the stringent bound on un-boosted $S_B-$electron scattering in DM direct detections. This tangible approach will be explored in this paper.

%%%%%%%%%%%%%%%%%%%%%%%%%%%%%%%%%%%%%%%%%%%%%%%
\section{DM Interactions and transitions}
\label{sec:Model}
%%%%%%%%%%%%%%%%%%%%%%%%%%%%%%%%%%%%%%%%%%%%%%%

In this paper we consider a scalar DM model to interpret the XENON1T excess. In this model $S_A$ is dark neutral and $S_B$ is charged under possible dark symmetry in the hidden sector. The DM particle $S_B$ is intermediated by a new vector boson $X$ to interact with electron. $X$ is assumed to couple to SM charged leptons, and the effective couplings are taken as
\begin{eqnarray}
\mathcal{L}_{X} \supset   e X_\mu \sum_\ell \epsilon_\ell \bar{\ell}  \gamma^\mu  \ell\; .
\end{eqnarray}
$X$ is considered as a light vector boson in general and here we do not specify a scenario such as a kinetic
mixing portal or a new gauged U$(1)$. The dark charged $S_B$ couples to the $X$ boson via
\begin{eqnarray}
\mathcal{L}_{X} \supset &-& e_D^{} X_\mu J^\mu_{\mathrm{DM}} + e_D^2 X^\mu X_\mu S_B^\dagger S_B   \; ,
\label{SBX}
\end{eqnarray}
where $J^\mu_{\mathrm{DM}}$ is the charged current of scalar DM $S_B$ with
\begin{eqnarray}
J^\mu_{\mathrm{DM}} = i [S_B^\dagger (\partial^\mu  S_B)- (\partial^\mu  S_B^\dagger) S_B] \;. \nonumber
\end{eqnarray}
Here we assume that the $X$ particle has a mass $\gtrsim$ 14 MeV~\cite{NA64:2019auh} and predominantly decays into $e^+ e^-$.

We also assume a real dark field $\phi$ coupled to both $S_A$ and $S_B$ and it mediates the transition between $S_A$ and $S_B$. Besides the kinetic energy terms, the scalar Lagrangian is given by
\begin{eqnarray}
-\mathcal{L}_{scalar} &\supset&
 \frac{1}{2}m_\phi^2\phi^2+\lambda_4\phi^4 + \mu_{S_A} S_A S_A^\dagger \phi + \lambda_{S_A} S_A S_A^\dagger \phi^2 \nonumber \\
&+&  \mu_{S_B} S_B S_B^\dagger \phi + \lambda_{S_B} S_B S_B^\dagger \phi^2   \; ,
\end{eqnarray}
The parameters $\mu_{S}\equiv \mu_{S_A} = \mu_{S_B}$ and $\lambda_{S}\equiv \lambda_{S_A} = \lambda_{S_B}$ are adopted. Here we assume $m_{\phi} > m_{S_A} \simeq m_{S_B} > m_X$ for simplicity and a small mass splitting $\Delta$ between $S_A$ and $S_B$ is introduced, i.e. $\Delta = m_{S_A} - m_{S_B} > 0$. To avoid the overabundance of $\phi$ in the early universe, we adopt $m_{\phi} > 2 m_X$ and thus the decay mode $\phi \to X X$ is generated at loop level from the $\phi S_B S_B^\dagger$ coupling and the charged current of $S_B$. For the case of $m_{S_B} > m_X$, the relic fraction of $S_B$ can be significantly depleted when the on-shell annihilation mode $S_B S_B^\dagger \to X X$ being opened. In addition, possible $\phi-$SM Higgs mixing is neglected here for simplicity (for the mixing case, the mixing with a rough upper limit of $\sin^2\theta \ll 10^{-3}$ can be allowed by experiments~\cite{Jia:2016pbe}). We should keep in mind that there may be more particles in the new sector and here we only consider the particles playing key roles in transitions between the SM and the dark sector.

To induce the keV electron scattering events via boosted $S_B$, the dark annihilation process of $S_A S_A^\dagger \to \phi $ $ \to S_B S_B^\dagger$ is considered to be dominant in $S_A S_A^\dagger$ annihilation. The annihilation cross section is given by
\begin{eqnarray}
\sigma_0 v_r \simeq \frac{\mu_S^4}{32 \pi m_{S_A}^2}   \frac{ \beta_f }{(s - m_\phi^2)^2 + m_\phi^2 \Gamma_{\phi}^2} \; ,
\label{eq:sigmav0}
\end{eqnarray}
where $v_r$ is the relative velocity and $s$ is the squared total invariant mass. In the non-relativistic limit, the phase space factor $\beta_f$ is
\begin{eqnarray}
\beta_f = \sqrt{1- \frac{4m_{S_B}^2}{s} } \approx \sqrt{\frac{v_r^2}{4} + \frac{2 \Delta}{m_{S_A}} }  ~ .
\end{eqnarray}
The decay width of $\phi$ is
\begin{eqnarray}
\Gamma_\phi = { \mu_X^2\over 8\pi m_\phi}\sqrt{1-{4m_X^2 \over m_\phi^2}} \Big({m_\phi^4 \over 4m_X^4}-{m_\phi^2 \over m_X^2}+3\Big) \; .
\end{eqnarray}
The annihilation of dark charged DM $S_B$ is mainly governed by $S_B S_B^\dagger \to X X$ via the $S_B-X$ coupling in Eq.~(\ref{SBX}). The annihilation cross section is
\begin{eqnarray}
\sigma_1 v_r &\simeq& {e_D^4 \sqrt{1-  {4 m_X^2 / s} } \over 16\pi m_{S_B}^2} \frac{8 m_{S_B}^4 - 8 m_{S_B}^2 m_X^2 +3 m_X^4}{( 2m_{S_B}^2 -m_X^2)^2}  \; .
\end{eqnarray}
The $p-$wave process $S_B S_B^\dagger \to X \to e^+ e^-$ is suppressed by $\epsilon_e^2$ and is negligible compared with the above annihilation process (see Ref.~\cite{Jia:2016uxs} for this $p-$wave dominant case).

%%%%%%%%%%%%%%%%%%%%%%%%%%%%%%%%%%%%%%%%%%%%%%%
\section{Boosted DM for the XENON1T excess}
\label{sec:GCE}
%%%%%%%%%%%%%%%%%%%%%%%%%%%%%%%%%%%%%%%%%%%%%%%

Assuming that the main component of DM is $S_A$ which has an NFW profile, the boosted DM particles $S_B$ can be produced by the present dark annihilation process of $S_A S_A^\dagger$ $\to S_B S_B^\dagger$. To obtain the benchmark velocity $v_b \sim$ 0.06 in Ref.~\cite{Fornal:2020npv} for boosted $S_B$, the value of $\Delta / m_{S_A}$ is required to be $\sim 0.0018$ as $\Delta \simeq \frac{1}{2} m_{S_B} v_b^2 \approx \frac{1}{2} m_{S_A} v_b^2$. The flux of the boosted DM $S_B$ over the full sky can be written as \cite{Agashe:2014yua}
\begin{eqnarray}
\hspace{-3mm}\Phi_{\mathrm{BDM}}^{4 \pi} \! = \! 1.6 \! \times \! 10^{-4} \! \mathrm{cm}^{-2}\mathrm{s}^{-1} \! \Big(\!\frac{1  \mathrm{GeV}}{ m_{S_A}}\!\Big)^2 \! \! \frac{\langle \sigma_0 v_r \rangle_{0}}{5 \! \times \! 10^{-26} \mathrm{cm}^{3}\mathrm{s}^{-1} } ~ ,
\end{eqnarray} 
where $\langle \sigma_0 v_r \rangle_{0}$ is today's thermally averaged annihilation cross section of $S_A S_A^\dagger$ and is suppressed by the phase space factor $\beta_f$.
Together with the flux of boosted DM $S_B$ hitting the earth detectors, the number of signal events $N_{\mathrm{sig}}$ via boosted DM $S_B$-electron scattering is
\begin{eqnarray}
N_{\mathrm{sig}} \propto  \sigma_\mathrm{elec}  \times  \Phi_{\mathrm{BDM}}^{4 \pi}  ~ ,
\end{eqnarray}
where $\sigma_\mathrm{elec}$ is the boosted $S_B$-electron scattering cross section mediated by $X$ boson
\begin{eqnarray}
\sigma_\mathrm{elec} \simeq  \frac{4 \alpha e_D^2 \epsilon_e^2 \mu_{e S_B }^2 }{m_X^4} ~ ,
\label{scatt-cs}
\end{eqnarray}
with $\mu_{e S_B }$ being the reduced mass of $m_e$ and $m_{S_B }$.

The signal events $N_{\mathrm{sig}}$ observed by XENON1T is about 40$-$70 events. In this case, the required scattering cross section $\sigma_\mathrm{elec}$ is~\cite{Fornal:2020npv}
\begin{eqnarray}
\sigma_\mathrm{elec} = 2.1 \times 10^{-31}  \mathrm{cm}^2  \Big( \frac{10^{-4} \mathrm{cm}^{-2}\mathrm{s}^{-1}}{\Phi_{\mathrm{BDM}}^{4 \pi}} \Big) \Big( \frac{N_{\mathrm{sig}}}{70} \Big) ~ .    \label{required-cs}
\end{eqnarray}
In order to obtain a large cross section in Eq.~(\ref{scatt-cs}), the mass $m_X$ (the parameter $\epsilon_e$) should be as small (large) as possible.
Substituting the mediator's mass $m_X = 17$ MeV, $e_D = 1$ and $\epsilon_e \lesssim 10^{-3}$ into Eq. (\ref{scatt-cs}), one finds the scattering cross section $\sigma_\mathrm{elec}\lesssim 10^{-35} \ \mathrm{cm}^2$ for $m_{S_B}\gg m_e$, and this value is smaller than the scattering cross section required by Eq. (\ref{required-cs}) even when DM mass as light as $\gtrsim$ 20 MeV. Thus, there is no enough boosted $S_B$ flux to produce the XENON1T excess for ordinary annihilations of $S_A$. However, if today's dark annihilation of $S_A$ is enhanced, the result is different. We consider the annihilation of $S_A$ is close to the $\phi$ resonance with the mass $2 m_{S_A}$ slightly above $m_\phi$. In this case today's dark annihilation $\langle \sigma_0 v_r \rangle_{0}$ will be significantly enhanced and can produce a large flux of boosted $S_B S_B^\dagger$. Moreover, for DM mass $\lesssim$ 10 MeV, the scattering cross section required by Eq. (\ref{required-cs}) can be significantly lowered, while such light DM particles will be in tension with constraints from BBN and CMB.

\begin{figure}[htbp!]
\includegraphics[width=0.45\textwidth]{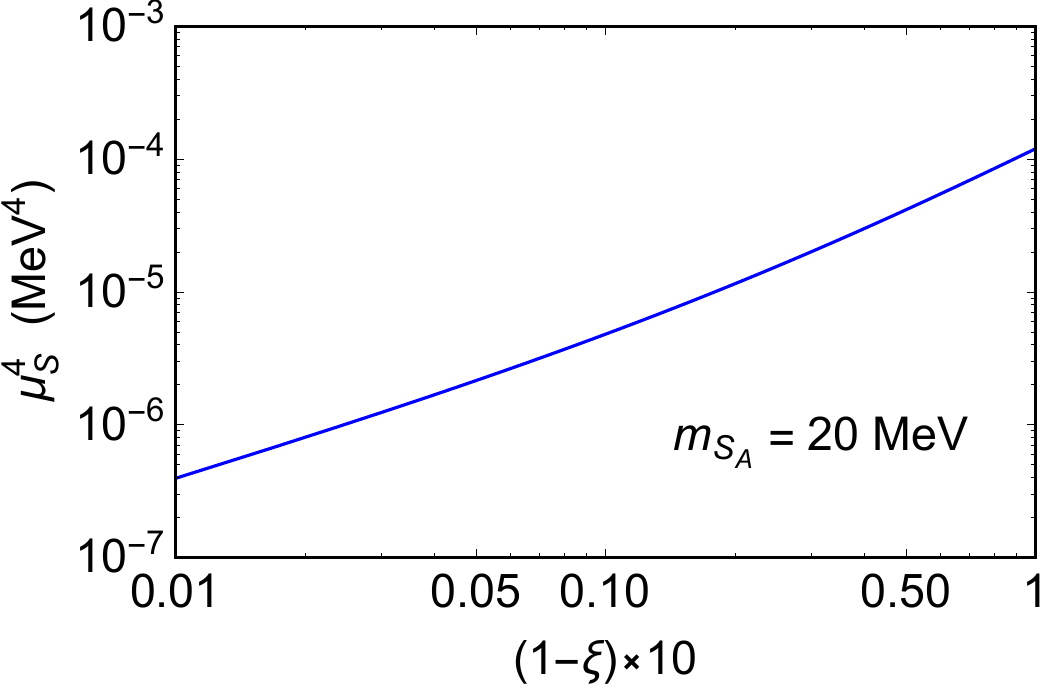} \vspace*{-1ex}
\caption{The coupling parameter $\mu_S^4$ as a function of $1 - \xi$ for $m_{S_A} =$ 20 MeV. Here the relic density of $S_A S_A^\dagger$ equal to 0.120 is adopted.}
\label{coup}
\end{figure}

\begin{figure}[htbp!]
\includegraphics[width=0.45\textwidth]{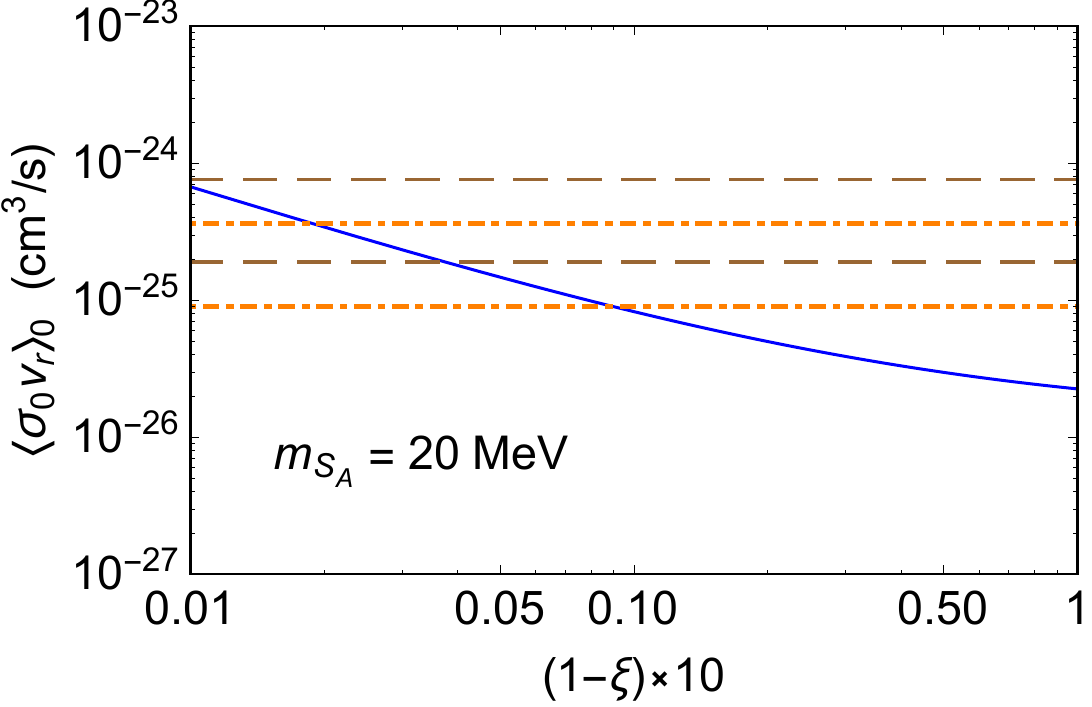} \vspace*{-1ex}
\caption{The relation between $\langle \sigma_0 v_r \rangle_{0}$ (solid curve) and $1 - \xi$ for $m_{S_A} =$ 20 MeV. The dot-dashed and dashed curves are for the case of $m_X$ = 17 MeV and 19 MeV, respectively. The lower (upper) limit of the dot-dashed and dashed curves corresponds to the annihilation cross sections required by the XENON1T excess for $e_D=1~(0.5)$.}
\label{cs-tod}
\end{figure}

Now we introduce a parameter $\xi \equiv m_\phi / 2 m_{S_A}$ in which $\xi$ is slightly smaller than 1. The cross section $\langle \sigma_0 v_r \rangle_{0}$ is sensitive to the value of $1 - \xi$. The decay width is negligible in DM annihilations when $1 - \xi \gg \Gamma_{\phi} / 4 m_\phi$ is satisfied. In the early Universe, DM chemically decouples from the thermal bath when the reaction rate $\Gamma(n \langle \sigma v_r \rangle)$ of DM particles drops below the Hubble expansion rate $H$. For the case of $\langle \sigma_1 v_r \rangle \gg \langle \sigma_0 v_r \rangle$ considered here, the DM $S_B$ freeze-out is a little later compared with the DM $S_A$. As the mass difference between $S_A$ and $S_B$ is very small, one has the number density $n_\mathrm{S_B} \simeq n_\mathrm{S_A}^\mathrm{eq}$ during the freeze-out period of $S_A$. Considering contributions from the $S_B S_B^\dagger \to S_A S_A^\dagger$ transition, the effective annihilation cross section of DM $S_A$ is equivalent to $2 \times \sigma_0 v_r$ during the freeze-out period (see the Appendix~\ref{s-a-f} for details). The relic density of DM is set by the annihilation cross section, and it can be evaluated using the general method without $s$-wave approximation \cite{Gondolo:1990dk,Kolb:1990vq,Griest:1990kh}. The coupling parameter $\mu_S$ as a function of $1 - \xi$ is derived with the relic density of $S_A S_A^\dagger$ nearly equal to the total DM relic density $\Omega_D h^2 =0.120 \pm 0.001$~\cite{Planck:2018vyg}, as shown in Fig.~\ref{coup}. Today's $S_A S_A^\dagger$ annihilation $\langle \sigma_0 v_r \rangle_{0}$ as a function of $1 - \xi$ is shown in Fig.~\ref{cs-tod}, with the value of $1 - \xi$ varying in a range of $10^{-3} - 10^{-1}$. Note that the relative velocity $v_r$ in the Galaxy is $\sim 10^{-3}$, and the $\langle \sigma_0 v_r \rangle_{0}$ is insensitive to $v_r$ in $s - m_\phi^2$ or the phase space factor $\beta_f$ in $\sigma_0 v_r$ in Eq.~(\ref{eq:sigmav0}) for the range of $1 - \xi$ of concern. In Fig.~\ref{cs-tod}, the solid curve is the corresponding $\langle \sigma_0 v_r \rangle_{0}$ for a given $1 - \xi$, and one can see that the annihilation is enhanced when $\xi$ is very close to 1. The dot-dashed and dashed curves are the annihilation cross sections required by the XENON1T excess for two benchmark values of [$m_X$ = 17 MeV, $\epsilon_e = 1\times10^{-3}$] and [$m_X$ = 19 MeV, $\epsilon_e = 0.85 \times10^{-3}$] adopted here, respectively. One can see that the resonance enhanced dark annihilation today can produce large boosted $S_B$ flux to account for the XENON1T excess.

\begin{figure}[htbp!]
\includegraphics[width=0.45\textwidth]{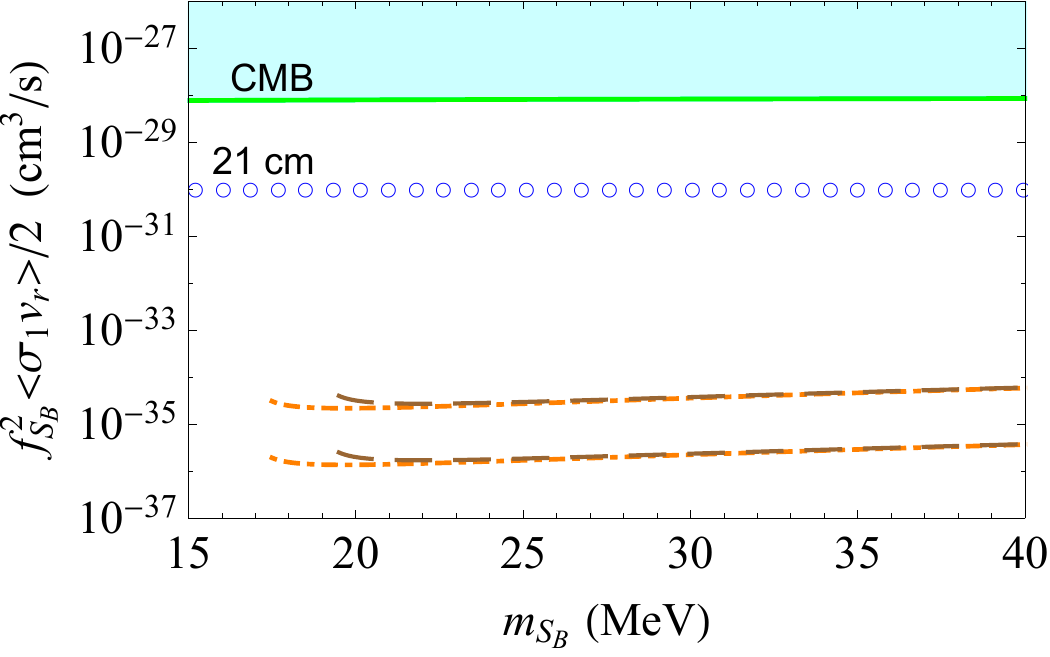} \vspace*{-1ex}
\caption{The revised annihilation cross section $f_{S_B}^2 \langle \sigma_1 v_r \rangle /2$ as a function of $m_{S_B}$. The dot-dashed and dashed curves are the revised annihilation cross sections for $m_X$ = 17 MeV and 19 MeV, respectively. The lower (upper) limit of the dot-dashed and dashed curves corresponds to the annihilation cross sections for $e_D=1~(0.5)$. The solid curve is the constraint from CMB~\cite{Slatyer:2015jla} and the empty dotted curve is the typical upper limit set by the anomalous 21-cm absorption with $T_m <$ 4 K at $z$ = 17.2~\cite{Liu:2018uzy}.
}
\label{ann1-cs}
\end{figure}

Next we give a brief discussion about the annihilation of $S_B$. The $S_B$ only contributes to a very small fraction $f_{S_B}$ of the total DM relic density, and the relic fraction $f_{S_B}$ (both $S_B$ and $S_B^\dagger$ included) can be obtained by the relation $f_{S_B} \simeq 4.4 \times 10^{-26}$ cm$^3 s^{-1}$/ $\langle \sigma_1 v_r \rangle$ ~\cite{Steigman:2012nb,Jia:2020icz}. The annihilation of $S_B S_B^\dagger$ is suppressed by the factor $f_{S_B}^2$ in observations, and revised annihilation cross section $f_{S_B}^2 \langle \sigma_1 v_r \rangle /2 $ is shown in Fig.~\ref{ann1-cs} with $e_D = 1$ and 0.5. For two benchmark values of $m_X$ = 17 MeV and 19 MeV, the dark matter masses $m_{S_B}$ ($m_{S_A}$) in a range of $m_{S_B}$ ($m_{S_A}$) $>$ $m_X$ are allowed by the constraint from CMB~\cite{Slatyer:2015jla} and the typical upper limit set by the anomalous 21-cm absorption~\cite{Liu:2018uzy}, as depicted in Fig.~\ref{ann1-cs}. Given $m_{S_A} =$ 20 MeV and $e_D$ = 1, $f_{S_B}$ is of order $\sim 10^{-10}$ for $m_X \approx$ 17$-$19 MeV.

%%%%%%%%%%%%%%%%%%%%%%%%%%%%%%%
\section{Direct detection of un-boosted DM}
\label{sec:DD}
%%%%%%%%%%%%%%%%%%%%%%%%%%%%%%%

Besides the boosted $S_B$ accounting for the XENON1T keV excess, there are a large amount of $S_A$ and $S_B$ with a regular velocity distribution around the earth. Now we turn to the un-boosted DM-electron scattering. First, we consider the un-boosted $S_A$. The $S_A -$electron scattering occurs at two-loop level from the $\phi- X X$ transition and $X-$electron coupling. The scattering cross section is far below the neutrino floor~\cite{Billard:2013qya,Mei:2017etc} in DM direct detections.

\begin{figure}[htbp!]
\includegraphics[width=0.45\textwidth]{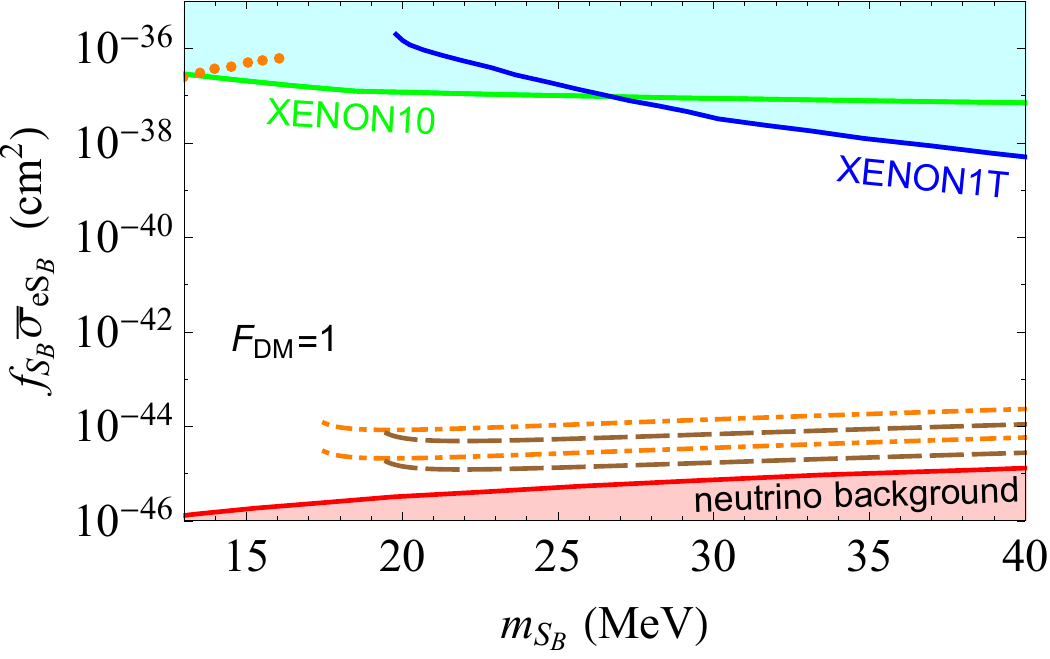} \vspace*{-1ex}
\caption{The effective scattering cross section $f_{S_B} \bar{\sigma}_{\rm e S_B}^{}$ as a function of $m_{S_B}$ for un-boosted $S_B S_B^\dagger$ in DM direct detections. The dot-dashed and dashed curves are the effective scattering cross section for the benchmark values as labeled in Figs.~\ref{cs-tod} and \ref{ann1-cs}. The lower (upper) limit of the dot-dashed and dashed curves corresponds to the cross sections for $e_D=1~(0.5)$. The upper two solid curves are the upper limits from XENON10~\cite{Essig:2017kqs} and XENON1T~\cite{XENON:2019gfn}. For comparison, the dotted curve on the top left is for the case of $m_{S_B} < m_X=16.7$ MeV~\cite{Jia:2016uxs}. The lower solid curve is the neutrino background~\cite{Billard:2013qya}.}\label{sb-electron}
\end{figure}

For the un-boosted $S_B$, the $S_B -$electron scattering is mainly contributed by the tree level process mediated by $X$ boson. The corresponding scattering cross section is
\begin{eqnarray}
\bar{\sigma}_{\rm e S_B}^{} \simeq  \frac{4 \alpha e_D^2 \epsilon_e^2 \mu_{e S_B }^2 }{m_X^4} ~ , \label{unboost-scatt}
\end{eqnarray}
with the form factor $F_\mathrm{DM}(q) = 1$~\cite{Essig:2011nj}. For $m_X=$ 17 MeV, $e_D=1$ and $\epsilon_e = 10^{-3}$, the scattering cross section is $\simeq 10^{-35} \mathrm{cm}^2$ which is above the bound set by XENON10/100 \cite{Essig:2017kqs,XENON10:2011prx}. This is why the ordinary interpretation of the XENON1T keV excess via DM-electron scattering seems in tension with DM direct detections, even though DM mass is as low as $\mathcal{O}(10)$ MeV. In this paper, the DM particles $S_B S_B^\dagger$ only make up a very small fraction $f_{S_B}$ of the total DM, and thus the tension can be relaxed due to the effective scattering cross section being $f_{S_B} \bar{\sigma}_{\rm e S_B}$ in DM direct detections. The result of the effective scattering cross section $f_{S_B} \bar{\sigma}_{\rm e S_B}$ is shown in Fig.~\ref{sb-electron}. One can see that the benchmarks we consider above can evade the constraints from CMB, 21 cm absorption and DM direct detections.

Moreover, for the case of $m_{S_B} < m_X$, enough boosted $S_B$ can also be produced and induces the XENON1T keV excess. The mass $m_{S_B}$ should be $\lesssim$ 13 MeV given the constraints from DM direct detections as shown by the dotted curve in Fig.~\ref{sb-electron}. Meanwhile, for $S_A$ and $S_B$, the mass $m_{S_B}$ ($m_{S_A}$) should be $\gtrsim$ 13 MeV with the bound from BBN~\cite{Ho:2012ug}. Thus, there is very little parameter space left for the case of $m_{S_B} < m_X$ when interpreting the XENON1T keV excess, and this case is not of our concern.

%%%%%%%%%%%%%%%%%%%%%%%%%%%%%%%%%%%%%%%%%%%%%%%
\section{Conclusion}
\label{sec:Con}
%%%%%%%%%%%%%%%%%%%%%%%%%%%%%%%%%%%%%%%%%%%%%%%

The boosted DM with a high speed about 0.05-0.1c and with a large DM-electron scattering cross section
(as large as $10^{-29}$ cm$^2$~\cite{Fornal:2020npv}) can interpret the XENON1T electron-event anomaly, as discussed in Refs.~\cite{Kannike:2020agf,Fornal:2020npv}.
However, can the boosted DM with such a large DM-electron scattering cross section be compatible with the present stringent bounds,
such as the BBN, low energy experiments, and DM direct detections?
This key question should be answered when proposing a model to explain the XENON1T anomaly. In this paper, we try to answer this question.
The proposed GeV DM has a large scattering cross section ($10^{-29}$ cm$^2$) between DM and electron~\cite{Fornal:2020npv}. The required mediator mass is as light as 0.1 MeV.
Such light mediator is excluded by the BBN which sets a lower mass bound on new thermal equilibrium particles, that is the mass of a new particle should be
above 10 MeV. Considering the constraints from BBN and low energy experiments, we derive that the scattering cross section between DM and electron is smaller in reality, roughly smaller than $10^{-35}$ cm$^2$. We find that light DM in MeV scale with an enhanced annihilation
source and a scattering cross section of $10^{-35}$ cm$^2$ can produce enough keV electron excess events observed by XENON1T and be allowed by the present
DM direct detections.

We have investigated an interpretation of the XENON1T excess by two scalar DM particles $S_A$ and $S_B$. The $S_A$ is neutral and $S_B$ is dark charged in the hidden sector. The boosted $S_B$ can be produced by the annihilation $S_A S_A^\dagger\to \phi \to S_B S_B^\dagger$ mediated by a scalar $\phi$. The $S_B-$electron scattering is intermediated by a vector boson $X$. The case of $m_{\phi} > m_{S_A} \simeq m_{S_B} > m_X$ is of our concern. Although the constraints from BBN, CMB and low-energy experiments require the boosted $S_B-$electron scattering cross section mediated by $X$ to be $\lesssim 10^{-35}~\mathrm{cm}^2$, MeV scale DM with a resonance enhanced dark annihilation today can still produce enough boosted DM and induce the XENON1T keV electron excess. The relic density of $S_B$ can be significantly reduced by the $s-$wave process of $S_B S_B^\dagger \to X X$, and thus this $s-$wave annihilation is allowed by the constraints from CMB and 21-cm absorption. A very small relic fraction of $S_B$ is compatible with the stringent bound on un-boosted $S_B$-electron scattering in DM direct detections. The $S_A$-electron scattering occurs at loop level and the scattering cross section is below the neutrino floor in direct detections.
We look forward to the further investigation of MeV DM and the corresponding new interactions in the future.

%%%%%%%%%%%%%%%%%%%%%%%%
\acknowledgments
%%%%%%%%%%%%%%%%%%%%%%%%

The work of T.~Li was supported by the National Natural Science Foundation of China (Grant No. 11975129, 12035008) and ``the Fundamental Research Funds for the Central Universities'', Nankai University (Grant No. 63196013). L.-B. Jia acknowledges support from the Longshan academic talent research supporting program of SWUST under Contract No. 18LZX415.

\appendix*

\section{The freeze-out of $S_A$}
\label{s-a-f}

In the case of $\langle \sigma_1 v_r \rangle \gg \langle \sigma_0 v_r \rangle$, the freeze-out of DM $S_A$ occurs first, and the DM $S_B$ decouples from the thermal bath later. For the DM particle $S_A$, the evolution of the number density $n_\mathrm{S_A}$ is given by
\begin{eqnarray}
\hspace{-1mm}\frac{d n_\mathrm{S_A}}{dt} \! + \! 3 n_\mathrm{S_A} \! H \!&=& \! - \langle \sigma_0 v_r \rangle_{\!S_A S_A^\dagger \!\to  S_B S_B^\dagger} (n_\mathrm{S_A}^2 \!\!-\! (n_\mathrm{S_A}^\mathrm{eq})^2)   \label{app1} \\
&&  +  \langle \sigma_0 v_r \rangle_{\!S_B S_B^\dagger \!\to S_A S_A^\dagger} (n_\mathrm{S_B}^2 \!\!-\! \frac{(n_\mathrm{S_B}^\mathrm{eq})^2}{(n_\mathrm{S_A}^\mathrm{eq})^2}n_\mathrm{S_A}^2) , \nonumber 
\end{eqnarray}
where $n_\mathrm{S_A}^\mathrm{eq}$ and $n_\mathrm{S_B}^\mathrm{eq}$ are the equilibrium number densities, and $H$ is the Hubble parameter. The DM particle $S_B$ is still in the thermal equilibrium with the thermal bath during the freeze-out period of $S_A$, and in this case, one has the number density $n_\mathrm{S_B} = n_\mathrm{S_B}^\mathrm{eq} \simeq n_\mathrm{S_A}^\mathrm{eq}$ with ignorable mass difference between $S_A$ and $S_B$. Thus, Eq. (\ref{app1}) can be rewritten as
\begin{eqnarray}
\hspace{-8mm}\frac{d n_\mathrm{S_A}}{dt} \!+\! 3 n_\mathrm{S_A}\! H\! \simeq \!- 2 \langle \sigma_0 v_r \rangle_{\!S_A S_A^\dagger \!\to S_B S_B^\dagger} \!(n_\mathrm{S_A}^2 \!\!-\! (n_\mathrm{S_A}^\mathrm{eq})^2) . \label{app2}
\end{eqnarray}
For DM $S_A$, the effective annihilation cross section is equivalent to $2 \times \sigma_0 v_r$ during the freeze-out period with the $S_B S_B^\dagger \to S_A S_A^\dagger$ transition considered.

%%%%%%%%%%%%%%%%%%%%%%%%%%%%%%%%%%%%%

\end{document}